# Testing the CCC+TL cosmology with cosmic-chronometer measurements of the Hubble parameter

**Rajendra P. Gupta***

*Department of Physics, University of Ottawa, Ottawa, Canada K1N 6N5*

**ABSTRACT**

In a recent paper, it was shown that the Covarying Coupling Constants and Tired Light (CCC+TL) hybrid model yields the Hubble parameter $H(z)$ that is substantially different from its measured value using differential aging of quiescent galaxies as cosmic chronometers (CC). It was claimed that the fit of the CCC+TL model to the $H(z)$ data results in a best-fit value for the parameter α, defining the strength of the co-variation of the constants, disagreeing with that for the SN Ia data at the $\sim 6\sigma$ level. In this paper we re-examine the assumptions underlying such a comparison. Cosmic-chronometer measurements are designed to be independent of cosmological priors, but they nevertheless rely on stellar population synthesis models, isochrones, and age-dating calibrations developed within standard stellar-evolution physics. Therefore, even before introducing any specific correction factor, the present CC compilation cannot be regarded as a model-independent falsification of CCC+TL without recomputing the relevant stellar population models in that framework. In the absence of such a recalculation, we ask a more limited question: what type and magnitude of modification to the effective differential-age relation would be sufficient to remove the claimed tension? We show that a phenomenological factor of the form $\sim(1+z_t)^{-3}$ with $z_t$ being the TL contribution to the observed redshift, motivated by the scaling of gas cooling times for galaxy formation in the CCC+TL framework compared to ΛCDM, is sufficient to reduce the apparent discrepancy in α to $\sim 0.13\sigma$. Since $z_t = 0$ for the stellar model primarily developed from local stellar observations, the stellar-aging methods may be unable to verify $\sim(1+z_t)^{-3}$ dependence.

## 1 INTRODUCTION

One of the fundamental goals of modern cosmology is to determine the expansion history of the Universe, quantified by the Hubble parameter H(z), as a function of redshift. Accurate and independent measurements of H(z) are crucial for constraining the cosmological parameters that govern the dynamics of the Universe and for testing the validity of the standard ΛCDM model and its extensions. Traditionally, cosmological observables such as Type Ia supernovae (SNe Ia; Riess et al. 1998; Perlmutter et al. 1999) and baryon acoustic oscillations (BAO; Eisenstein et al. 2005; Cole et al. 2005) have been used to trace the expansion history. However, these methods rely on integrated quantities, such as luminosity and angular diameter distances, which depend on the integral of $H(z)^{-1}$over redshift, thereby introducing degeneracies among cosmological parameters.

An alternative and complementary approach is provided by the cosmic chronometer (CC) method, first introduced by Jimenez & Loeb (2002). The technique enables direct, differential measurements of the Hubble parameter, independent of any cosmological model, by exploiting the relation

$$H(z) = -\frac{1}{1+z}\frac{dz}{dt}, \tag{1}$$

which connects the rate of change of redshift with cosmic time. The differential age evolution $dt/dz$ can be determined observationally by measuring the relative age difference between passively evolving galaxies that formed their stellar populations at similar epochs. Since this method depends only on differential ages, not absolute ages, it avoids several systematic uncertainties that plague traditional cosmological probes.

Over the past two decades, the CC technique has been applied extensively using large spectroscopic surveys such as SDSS, BOSS, and VIPERS (e.g., Moresco et al. 2012, 2016, 2020; Stern et al. 2010; Ratsimbazafy et al. 2017), yielding precise $H(z)$ determinations over $0 < z \lesssim 2$. These results have been instrumental in constraining the dark energy equation of state, testing deviations from general relativity, and probing the so-called "Hubble tension" between local and early-Universe measurements of $H_0$ (Bernal, Verde & Riess 2016; Verde, Treu & Riess 2019). The spectroscopic surveys, including DESI (DESI Collaboration et al. 2016), Euclid (Laureijs et al. 2011), and the Roman Space Telescope (Spergel et al. 2015), have further enhanced the statistical precision and redshift coverage of CC-based measurements. Theoretically, Okamatsu and Takahashi (2025) have attempted to show that the effective field theory (EFT) of dark energy provides a model-independent framework for studying cosmology within the context of scalar-tensor theories.

*E-mail: rgupta4@uottawa.ca

Using the CC data, Lei et al. (2025) falsified the cosmological model that allows coupling constants to evolve in a correlated manner and accounts for the tired light effect contributing to the observed redshift in an expanding universe. This, covarying coupling constant plus tired light (CCC+TL) model has already been successful in alleviating the 'impossible early galaxy problem' and fitting the SNe Ia Pantheon+ data (Gupta 2023). Additionally, it is consistent with a) the BAO and CMB sound horizon observations (Gupta 2024a), b) galaxy formation time scales at cosmic dawn and time dilation (Gupta 2024b), c) galaxy rotation curves and galaxy cluster dynamics (2025a), d) mass, size, density, and luminosity evolution of galaxies (2025b), and e) gravitational lensing and DESI findings of increasing dark energy density with redshift (Gupta 2026).

Our main objective in this paper is to examine Lei et al.'s work (Lei et al. 2025) and explore whether the assumptions made in their paper are consistent with the CCC+TL model. In Eq. 1, $z$ and $dz$ can be determined with a very high precision. The problem arises in determining $dt$, as it must be cosmology-independent; i.e., stellar models and their derivatives, such as SED (Spectral Energy Distribution) and SPS (Stellar Population Synthesis) models, must not be constrained by parameters derived from any cosmological model. As Moresco et al. (2022) wrote: "A final, yet important point to keep in mind is that, in order to be cosmology-independent, the CC approach must rely on age estimates that do not assume any cosmological prior. This is a very important point, since in many (if not in most) analyses, a cosmologically-motivated upper prior on age is adopted in order to break or minimize the previously discussed degeneracies. Of course, for the CC method to be used as a test for cosmology, it is of paramount importance to obtain a robust age estimate without introducing any (prior) dependence on a cosmological model, in order to avoid circularity and, basically, retrieve the cosmological model used as a prior."

This paper is structured as follows: Section 2 presents the essentials of CCC+TL model relevant for this paper; in Section 3 we discuss the potential effect of the age of the Universe in determining $dt$; Section 4 is devoted to the methodology of analysing the CC data with CCC+TL, ΛCDM, CCC, and ΛCDM+TL models; Section 5 describes the results; Section 6 is used for discussion; and Section 7 provides the conclusion.

## 2 THE CCC+TL MODEL ESSENTIALS

We may write by equating the expression for the proper distance $d_p$ determined for the CCC and the TL effects (Gupta 2023, Eq. 44) at redshift $z$

$$d_p(z) = c\int_0^{z_c} \frac{dz}{(H_c+\alpha)(1+z)^{(3/2)}f(z)^{-(1/2)}-\alpha} = cH_t^{-1}\ln\left[\frac{1+z}{1+z_c}\right] \tag{2}$$

Here, $c$ is the speed of light, $H_c$ is the Hubble constant corresponding to CCC and $H_t$ to TL with $H_0 = H_c + H_t$, $(1+z) = (1+z_c)(1+z_t)$ with $z_c$ the CCC expanding Universe redshift and $z_t$ due to TL, and $f(t) = \exp[\alpha(t-t_0)]$ with $\alpha$ determining the strength of the CCC variation and $t_0$ the current cosmic time.

Equation (2) may be rewritten to express $z$ in terms of $z_c$:

$$(1+z) = (1+z_c)\exp\left(H_t\int_0^{z_c} \frac{dz}{(H_c+\alpha)(1+z)^{(3/2)}f(z)^{-(1/2)}-\alpha}\right) \tag{3}$$

Taking the time derivative:

$$\frac{dd_p}{cdt} = \frac{dz_c}{dt}\left((H_c+\alpha)(1+z_c)^{(3/2)}f(z_c)^{-(1/2)}-\alpha\right)^{-1}$$

$$= H_t^{-1}\frac{1+z_c}{1+z}\left[\frac{1}{1+z_c}\frac{dz}{dt} - \frac{1+z}{(1+z_c)^2}\frac{dz_c}{dt}\right]$$

$$= H_t^{-1}\left[\frac{1}{1+z}\frac{dz}{dt} - \frac{1}{1+z_c}\frac{dz_c}{dt}\right], \text{ or} \tag{4}$$

$$\frac{1}{1+z}\frac{dz}{dt} = \frac{dz_c}{dt}\left[H_t\left((H_c+\alpha)(1+z_c)^{(3/2)}f(z_c)^{-(1/2)} - \alpha\right)^{-1} + \frac{1}{1+z_c}\right], \text{ or} \tag{5}$$

$$\frac{1}{1+z}\frac{dz}{dt} = \frac{1}{1+z_c}\frac{dz_c}{dt}\left[1 + \frac{(1+z_c)H_t}{\left((H_c+\alpha)(1+z_c)^{(3/2)}f(z_c)^{-(1/2)}-\alpha\right)}\right]. \tag{6}$$

From (Gupta 2023)

$$H_t = \frac{(H_c+\alpha)}{2}\left(3+\frac{\alpha}{H_c}\right), \text{ and} \tag{7}$$

$$-\frac{1}{1+z_c}\frac{dz_c}{dt} = \frac{\dot{a}_c}{a_c} = -\alpha + (H_c+\alpha)f(z_c)^{-1/2}(1+z_c)^{3/2}. \tag{8}$$

Therefore,

$$-\frac{1}{1+z}\frac{dz}{dt} = (H_c+\alpha)f(z_c)^{-1/2}(1+z_c)^{3/2} - \alpha + \frac{(H_c+\alpha)}{2}\left(3+\frac{\alpha}{H_c}\right)(1+z_c). \tag{9}$$

The CCC function $f(z_c)$ is given by

$$f(z_c)^{-(1/2)} = \left(-\frac{D}{2A} + \left(\left(-\frac{D}{2A}\right)^2 + \left(\frac{C}{3A}\right)^3\right)^{1/2}\right)^{1/3} + \left(-\frac{D}{2A} - \left(\left(-\frac{D}{2A}\right)^2 + \left(\frac{C}{3A}\right)^3\right)^{1/2}\right)^{1/3}, \text{ with} \tag{10}$$

$$A = 1 - C,\ C = \frac{3}{2}\frac{(H_c+\alpha)}{\alpha}, D = -(1+z_c)^{(-3/2)}. \tag{11}$$

Similarly, we may write the corresponding expressions for the CCC, ΛCDM, and ΛCDM+TL, respectively, as follows:

$$-\frac{1}{1+z}\frac{dz}{dt} = (H_0 + \alpha)f(z)^{-1/2}(1+z)^{3/2} - \alpha, \quad (12)$$

$$-\frac{1}{1+z}\frac{dz}{dt} = H_0\left\{(1+z)^3\Omega_{m,0} + 1 + \Omega_{m,0}\right\}^{1/2}, \quad (13)$$

$$-\frac{1}{1+z}\frac{dz}{dt} = \left[\begin{matrix} H_l\left\{(1+z_l)^3\Omega_{m,0} + 1 + \Omega_{m,0}\right\}^{1/2} + \\ \frac{3}{2}\Omega_{m,0}H_l(1+z_l) \end{matrix}\right] \quad (14)$$

Here, $\Omega_{m,0}$ is the relative matter density, $z_l$ is the redshift and $H_l$ expanding universe contribution to $z$ and $H_0$, respectively, when considering ΛCDM and ΛCDM+TL models.

## 3 CC AND COSMOLOGY

Stellar population ages used in cosmic-chronometer analyses are not obtained by imposing the age–redshift relation of a particular cosmological model. Rather, they are inferred from stellar population synthesis models, stellar isochrones, spectral libraries, and assumptions regarding the IMF, metallicity and star-formation history (Bruzual & Charlot 2003; Maraston 2005; Conroy, Gunn & White 2009; Conroy 2013). This is the reason that stellar and globular-cluster ages have historically been used as independent cosmological probes. For example, in the late 1990s, globular-cluster ages appeared to exceed some estimates of the age of the Universe (Jimenez et al. 1997), and Krauss & Chaboyer (2003) argued that this tension provided evidence for a cosmological model with a dominant dark-energy component. More recently, age determinations have continued to provide useful constraints in discussions of the Hubble tension and early structure formation (Boylan-Kolchin & Weisz 2021; Cimatti et al. 2019; Jimenez, Verde, Treu & Stern 2019; López-Corredoira et al. 2024; Valcin et al. 2026).

In practical galaxy-evolution studies, cosmologically motivated upper-age priors may be introduced to break degeneracies in SED or full-spectrum fitting. However, this is not the procedure appropriate for cosmic chronometers. In CC applications, such priors must be removed, otherwise the inferred $H(z)$ would partly reconstruct the cosmology assumed in the prior. This point has been emphasized in the CC literature (Jimenez & Loeb 2002; Moresco 2015; Moresco et al. 2012, 2016, 2020, 2022; Jiao et al. 2023; Tomasetti et al. 2023). The CC method therefore relies on the differential age evolution of passively evolving galaxies, using relative age differences rather than absolute ages. The use of differential ages reduces sensitivity to modelling choices such as IMF, metallicity, and SPS library assumptions, although it does not eliminate all stellar-population systematics.

The fitted stellar age may be luminosity-weighted or mass-weighted, depending on the specific fitting method and stellar population model. Isochrones used in SPS modelling are theoretical stellar-evolution tracks constructed over a broad range of ages and metallicities, and are based on stellar-evolution physics such as nuclear reaction rates, opacities, diffusion, mixing length and mass loss. Commonly used isochrone families include Padova, MIST and BaSTI models (Pietrinferni et al. 2004; Dotter et al. 2008; Bressan et al. 2012; Dotter 2016; Choi et al. 2016; Marigo et al. 2017; Hidalgo et al. 2018). They are calibrated and tested using local stellar benchmarks and stellar systems, including the Sun, Galactic star clusters and globular clusters, but they are not constructed by imposing that all inferred stellar ages must lie below the cosmological age of the Universe.

This distinction is important for the present work. We do not claim that the published CC ages are directly invalidated by the longer CCC+TL cosmic timeline. The ages inferred for CC galaxies are determined from observed spectra and stellar population models, not from the assumed cosmological age–redshift relation. Instead, the issue is whether the same stellar-population clock can be used without modification when testing a cosmology in which the redshift decomposition and the dimensional constants differ from the standard framework. In CCC+TL, the observed redshift $z$ is decomposed as $(1+z) = (1+z_c)(1+z_t)$ with $z_c$ due to expansion and $z_t$ due to tired light, and the cosmic timeline differs substantially from that in ΛCDM (Gupta 2023, 2024a, 2024b). A fully self-consistent CC test would therefore require recomputing the relevant stellar evolution and SPS ingredients within the CCC+TL framework. Such a recalculation is beyond the scope of the present paper.

It follows from the preceding considerations that the current CC compilation cannot, by itself, be used as a model-independent falsification of CCC+TL, since the stellar-evolution and SPS ingredients entering the published CC measurements have not been recalculated within the CCC+TL framework. We therefore adopt a more limited and explicitly phenomenological approach. We ask whether a simple modification of the effective differential-age relation, motivated by the scaling of cooling times for galaxy formation in the CCC+TL framework compared to ΛCDM (Gupta 2024b), is sufficient to remove the discrepancy reported by Lei et al. (2025). This is not intended as a definitive derivation of the CC correction factor. Rather, it is a sufficiency test showing that the observed mismatch can be resolved by a relatively modest modification of the CC trend.

The CC approach measures the Hubble parameter $H(z)$ using the differential age evolution of passive galaxies (Jimenez & Loeb 2002; Moresco et al. 2012) as per Eq. (1). They argued that since $dt$ depends on relative rather than absolute galaxy ages, differential aging minimizes the effect of cosmological constraint; it plays only an indirect role. According to them, even if the absolute age scale were uncertain, the ratio $\Delta z/\Delta t$ should remain robust, provided consistent models are used across redshifts (see also Moresco 2015; Moresco et al. 2020).

In the cosmic-chronometer framework, the reconstructed stellar ages are not directly altered by adopting a cosmological model with a stretched global time scale. For high-redshift galaxies, age estimates are obtained by comparing observed spectral features—such as the continuum slope, absorption-line strengths, and spectral breaks—with synthetic stellar population spectra. These synthetic spectra are constructed from stellar isochrones, empirical or theoretical stellar libraries, and assumptions regarding the initial mass function (IMF), metallicity, and star-formation history (Conroy 2013). Thus, the age determination itself is based on stellar population modelling rather than on the cosmological age–redshift relation. The CC method then associates galaxies in each redshift bin, typically the oldest passively evolving systems, with their inferred stellar ages in order to trace the differential age–redshift relation. The resulting $dt/dz$ measurement is used to infer whether the observed expansion rate at a given redshift is consistent with, slower than, or faster than that predicted by a given cosmological model.

A possible scale for such a phenomenological modification is suggested by the CCC+TL treatment of galaxy formation timescales. In previous work, it was argued that the free-fall and cooling times relevant to early structure formation are modified when the tired-light contribution to the observed redshift is included (Gupta 2024b; see also Laursen 2023). Compared with the corresponding ΛCDM expression, the free-fall time scales as $[(1+z)/(1+z_c)]^{1.5}$, whereas the radiative cooling time scales as $[(1+z)/(1+z_c)]^3$, i.e., $(1+z_t)^3$. Since radiative cooling is a limiting process in the formation of the first collapsed stellar systems, this scaling provides a physically motivated timescale that may influence the effective relation between galaxy formation history and observed redshift.

However, this argument should not be overinterpreted. Free-fall and cooling times are quantities entering galaxy-formation modelling; they do not *directly* enter the SED or full-spectrum fitting procedure used to infer the ages of CC galaxies, but *indirectly* through the covarying coupling constants. The measured spectral ages, metallicities and star-formation histories are obtained from SPS modelling and from spectral diagnostics such as D4000 and absorption features (Bruzual & Charlot 2003; Maraston 2005; Conroy 2013; Moresco et al. 2012, 2016, 2020, 2022; Borghi, Moresco & Cimatti 2022). The role of the cooling-time scaling here is therefore not to provide a first-principles derivation of the transformation of CC ages, but to motivate a simple phenomenological form for the effective differential-age correction.

We therefore introduce $Q(z) \equiv [b(1+z_t)]^3$, where $b$ is a phenomenological parameter. This parameter absorbs the fact that CC galaxies are not first-generation stellar systems and that subsequent star formation, chemical enrichment, feedback and quenching can dilute the memory of the original formation timescale. The relevance of metallicity and star-formation history to CC age determinations has been emphasized in the CC literature, particularly in discussions of the D4000 calibration and SPS-systematic budget (Moresco et al. 2012, 2016, 2020, 2022; Jiao et al. 2023; Tomasetti et al. 2023). Thus, $Q(z)$ should be viewed as an effective phenomenological correction rather than as a direct transformation law for stellar ages.

We tested whether $b$ deviates significantly from unity by performing a three-parameter MCMC fit with $b \in [0.5, 3.0]$. The posterior values are statistically consistent with $b = 1$, supporting its use as a simple benchmark for the present analysis. This result, however, should not be interpreted as a derivation of the correction factor. A definitive determination of $Q(z)$ would require recalculating stellar evolution, isochrones and SPS predictions in the CCC+TL framework.

To compare the models with the tired-light component (CCC+TL and ΛCDM+TL) with observations, we need to modify Eqs. 9 and 14. In their expressions of $[-\{1/(1+z)\}dz/dt]$, $dt$ is stretched by $(1+z_t)^3$. Thus, we may write these equations for CCC+TL as follows:

$$-\frac{1}{1+z}\frac{dz}{dt} = \left[(H_c+\alpha)f(z_c)^{-1/2}(1+z_c)^{3/2} - \alpha + \frac{(H_c+\alpha)}{2}\left(3+\frac{\alpha}{H_c}\right)(1+z_c)\right](1+z_t)^3. \quad (15)$$

Similarly, for the ΛCDM+TL, we get

$$-\frac{1}{1+z}\frac{dz}{dt} = \left[H_l\left\{(1+z_l)^3\Omega_{m,0} + 1 + \Omega_{m,0}\right\}^{1/2} + \frac{3}{2}\Omega_{m,0}H_l(1+z_l)\right](1+z_t)^3. \quad (16)$$

The stellar models used for determining stellar ages have been developed based on observations in the local universe, i.e., at redshifts close to 0, and therefore, $z_t$ also close to 0. Such models will not notice the effect of $z_t$.

We will only compare CCC+TL and ΛCDM models as they are the most relevant for the CC comparison. In the following sections, we have added the subscript 0 to explicitly define the constants at the current time; for example $H_{c,0} \equiv H_c$.

## 4 METHODOLOGY OF ANALYSING CC DATA

*Observational Data*: We employ 32 measurements of the Hubble parameter $H(z) = -1/(1+z) \times dz/dt$ obtained via the cosmic chronometer technique, spanning the redshift range $0.07 \leq z \leq 1.965$. These measurements are drawn from the compilation of Moresco et al. (2022), comprising data from Simon, Verde & Jimenez (2005), Stern et al. (2010), Moresco et al. (2012), Zhang et al. (2014), Moresco (2015), Moresco et al. (2016),

Ratsimbazafy et al. (2017), and Borghi, Moresco & Cimatti (2022). The CC method provides cosmology-independent estimates of H(z) by measuring the differential age evolution of massive, passively evolving galaxies as a function of redshift (Jimenez & Loeb 2002).

*MCMC Parameter Estimation*: We determine the posterior distributions of the model parameters using the affine-invariant Markov Chain Monte Carlo (MCMC) ensemble sampler `emcee` (Foreman-Mackey et al. 2013). For both models, we adopt uniform priors: $H_0 \in [50, 100]$ km s⁻¹ Mpc⁻¹ and $\Omega_{m,0} \in [0.2, 0.5]$ for ΛCDM; $H_{c,0} \in [30, 80]$ km s⁻¹ Mpc⁻¹ and $\alpha \in [-0.9H_{c,0}\ -0.6H_{c,0}]$ for CCC+TL.

The Gaussian log-likelihood is

$$\ln L\ =\ -\frac{1}{2}\sum_i \frac{[H_{obs}(z_i)-H_{model}(z_i)]^2}{\sigma_i^2}, \quad (17)$$

where $H_{obs}(z_i)$ and $\sigma_i$ are the observed Hubble parameter values and their associated $1\sigma$ uncertainties from the CC data. We employ 32 walkers with 5000 steps each, discarding the first 40 per cent as burn-in, yielding 96,000 effective posterior samples per model. Convergence is assessed via the mean acceptance fraction, which falls within the recommended range of 0.2–0.8 for both models.

Table 1. MCMC fitting results for the ΛCDM and CCC+TL models using 32 cosmic chronometer $H(z)$ measurements. Quoted uncertainties are 68 per cent credible intervals (16th and 84th percentiles of the posterior distribution). The DIC is the Deviance Information Criterion and $p_D$ is the effective number of parameters.

| **Parameter** | **ΛCDM** | **CCC+TL** |
|---|---|---|
| $H_0$ [km s⁻¹ Mpc⁻¹] | 67.66 (+2.91 / −2.96) | — |
| $\Omega_{m,0}$ | 0.328 (+0.063 / −0.054) | — |
| $H_{c,0}$ [km s⁻¹ Mpc⁻¹] | — | 52.68 (+3.41 / −3.29) |
| α [km s⁻¹ Mpc⁻¹] | — | −42.32 (+3.82 / −4.00) |
| $H_{t,0}$ (derived) | — | 11.38 |
| $H_{0,tot}$ (derived) | — | 64.06 |
| $\chi^2$ / dof | 14.75 / 30 | 15.86 / 30 |
| $\chi^2_{red}$ | 0.492 | 0.529 |
| $p_D$ | 1.86 | 1.87 |
| DIC | 18.50 | 19.60 |

$\Delta DIC\ (\Lambda CDM - CCC + TL) = -1.10$, i.e., no significant preference for either model).

*Model Comparison*: To compare the relative goodness of fit of the two models, we employ the Deviance Information Criterion (DIC; Spiegelhalter et al. 2002), defined as

$$DIC\ =\ \overline{D}\ +\ p_D, \quad (18)$$

where $\overline{D} = \langle D(\theta)\rangle$ is the posterior mean of the deviance $D(\theta) = -2\ln L(\theta)$, and $p_D = \overline{D} - D(\bar{\theta})$ is the effective number of parameters, with $\bar{\theta}$ denoting the posterior mean parameter vector. The DIC generalizes the classical AIC by accounting for the posterior distribution rather than relying solely on the maximum-likelihood estimate. A lower DIC indicates a preferred model; differences $|\Delta DIC| < 2$ are considered inconclusive, 2–6 indicate moderate evidence, 6–10 strong evidence, and >10 very strong evidence (Spiegelhalter et al. 2002).

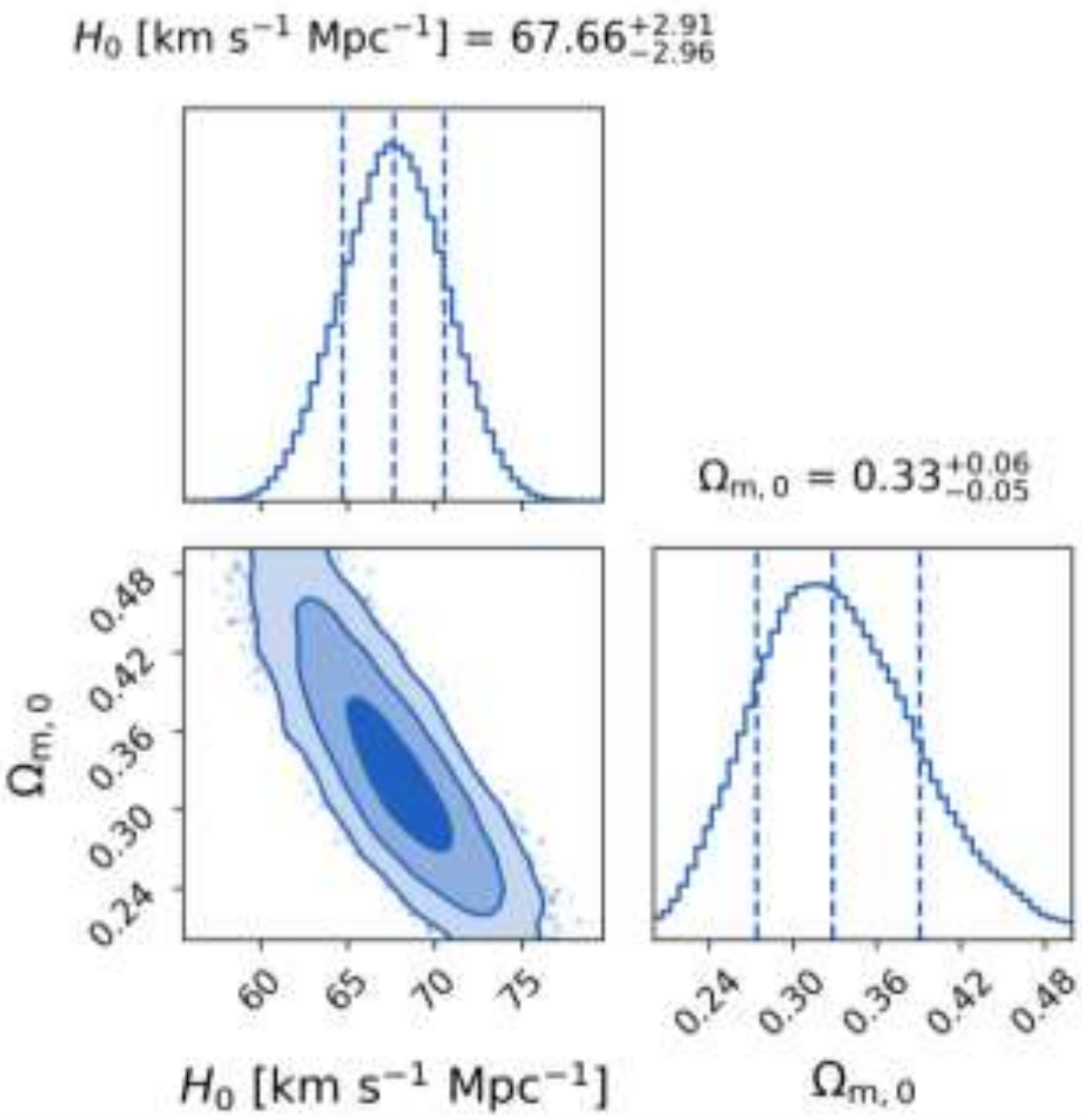

Figure 1. Posterior distribution for the ΛCDM model parameters $H_0$ and $\Omega_{m,0}$ from the MCMC analysis of 32 cosmic chronometer measurements. Contours show the 1σ, 1.5σ, and 2.33σ credible regions. Dashed lines indicate the 16th, 50th, and 84th percentiles.

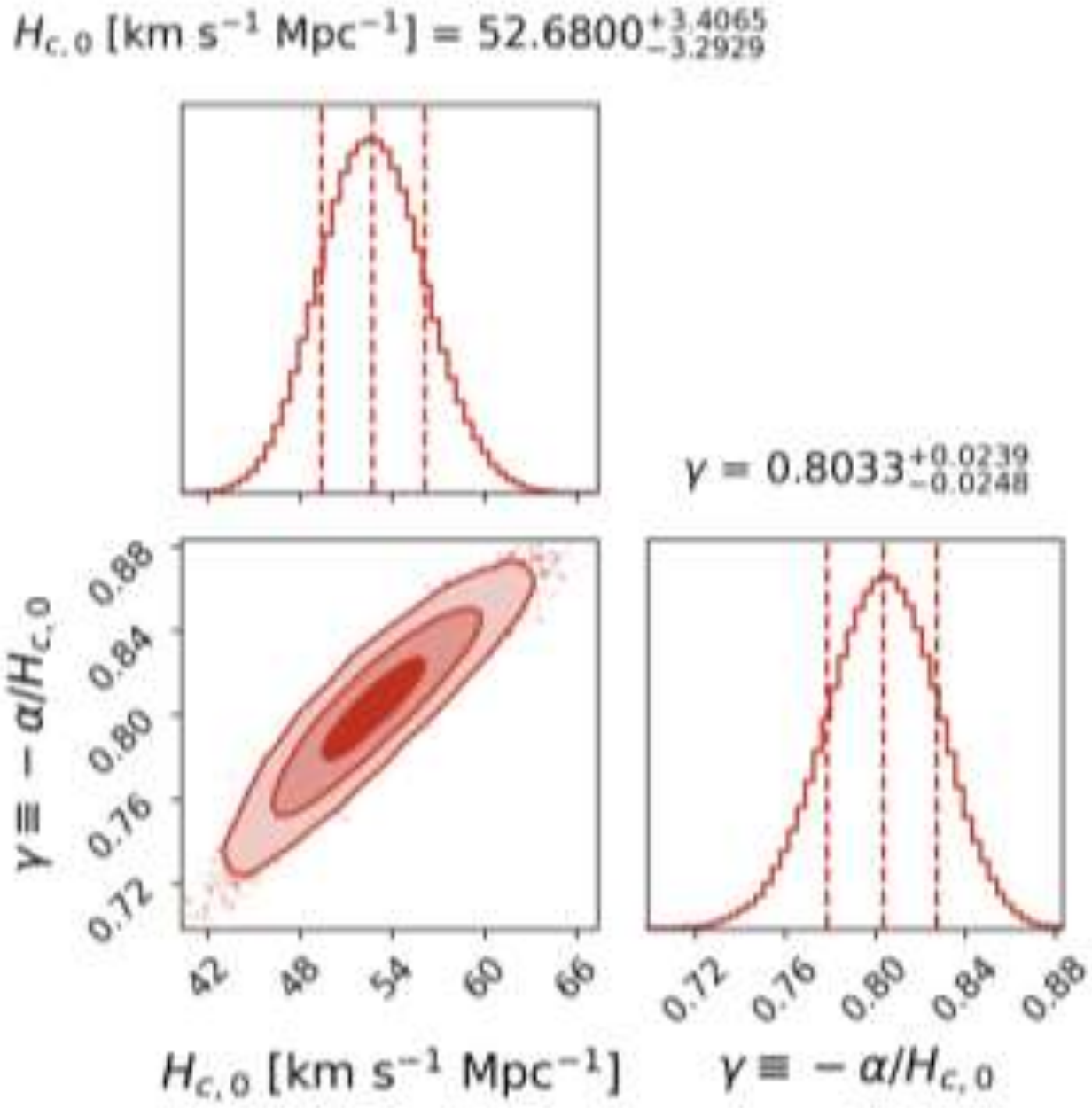

Figure 2. As Fig. 1, but for the CCC+TL model parameters $H_{c,0}$ and $\alpha$. As the two parameters are strongly correlated, we have used $\gamma \equiv -\alpha/H_{c,0}$ instead of $\alpha$ for the contours plot.

## 5 RESULTS

*Parameter Constraints:* Table 1 summarizes the best-fit parameter values (posterior medians with 68 percent credible intervals), goodness-of-fit statistics, and DIC values for both models. The posterior distributions are shown as corner plots in Figs 1 and 2, and the corresponding best-fit $H(z)$ curves are overlaid on the CC data in Fig. 3.

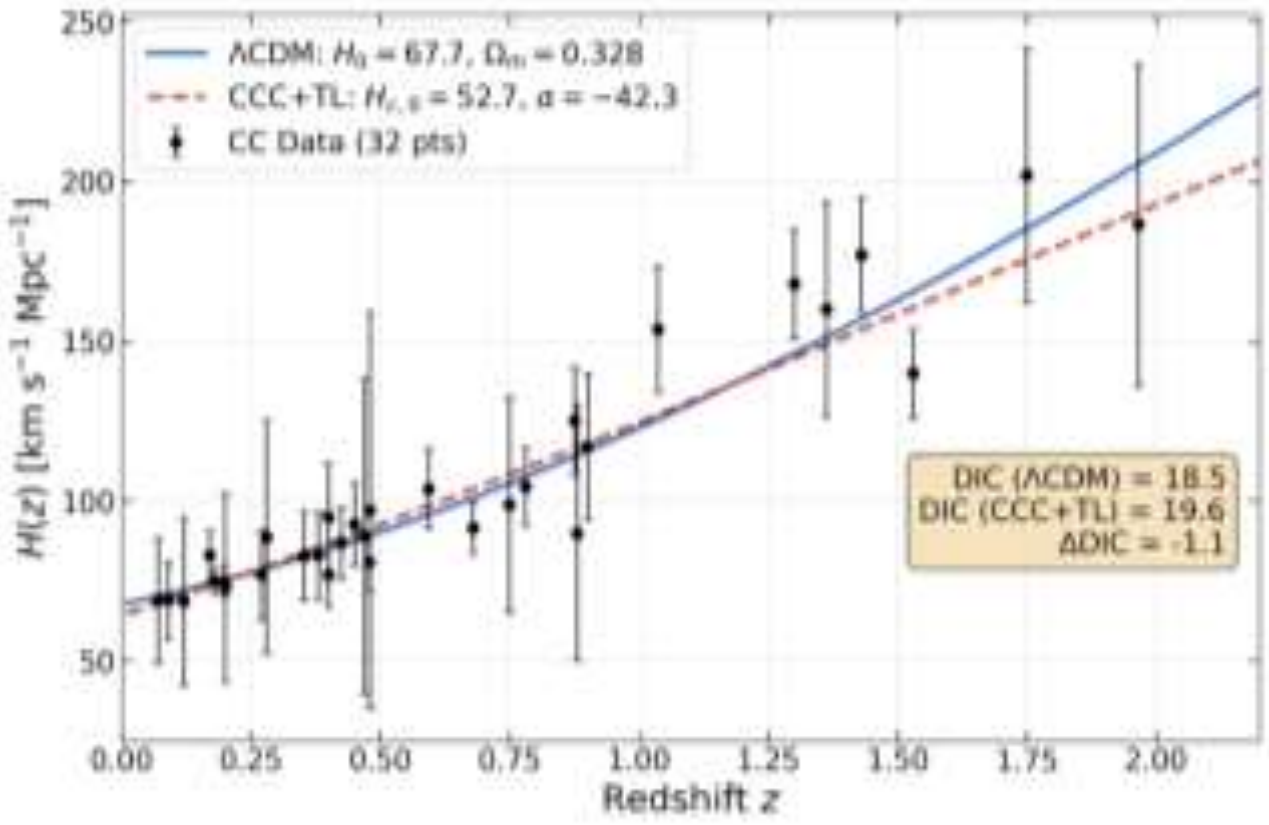

Figure 3. The 32 cosmic chronometer H(z) measurements (black points with 1σ error bars) compared with the best-fit ΛCDM (blue solid) and CCC+TL (red dashed) models. The DIC values for both models are annotated.

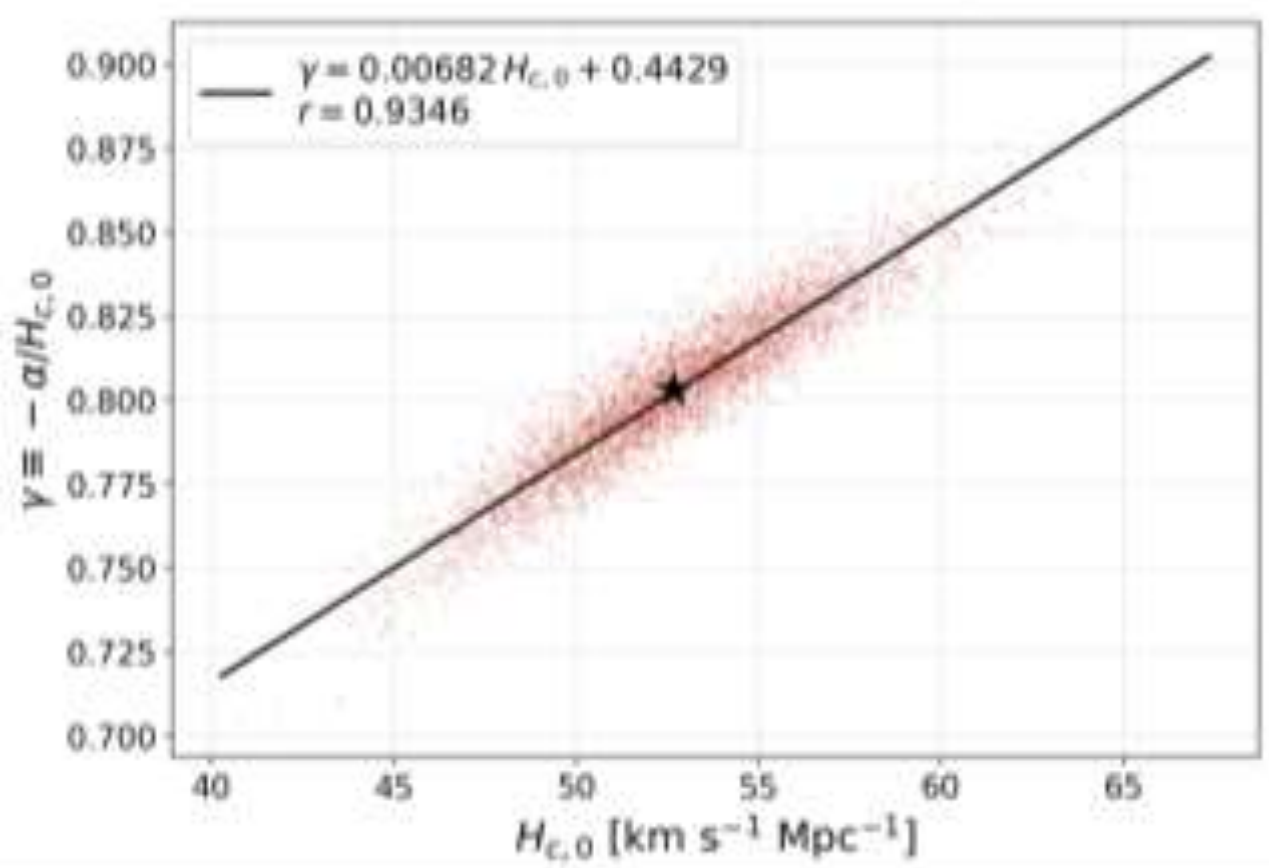

Figure 4. Scatter plot of posterior samples in the $(H_{c,0}, \alpha)$ plane for the CCC+TL model (red points), overlaid with the best-fit linear relation $\alpha = -1.1634 \times H_{c,0} + 18.92$ (black line). The star marks the posterior median. The near-unity Pearson correlation coefficient $|r| = 993350$ demonstrates that α is almost entirely determined by $H_{c,0}$.

For the ΛCDM model, we obtain $H_0 = 67.66^{+2.91}_{-2.96}$ km s⁻¹ Mpc⁻¹ and $\Omega_{m,0} = 0.328^{+0.063}_{-0.054}$, consistent with Planck 2018 results (Planck Collaboration VI 2020). The reduced chi-squared is $\chi^2/\nu = 14.75/30 = 0.49$, indicating a good fit with no evidence of systematic tension.

For the CCC+TL model, we find $H_{c,0} = 52.68^{+3.41}_{-3.29}$ km s⁻¹ Mpc⁻¹ and $\alpha = -42.32^{+3.82}_{-4.00}$ km s⁻¹ Mpc⁻¹. The derived tired-light component is $H_{t,0} = 11.38$ km s⁻¹ Mpc⁻¹, yielding a total Hubble constant $H_0 = 64.15$ km s⁻¹ Mpc⁻¹. The reduced chi-squared is $\chi^2/\nu = 15.86/30 = 0.53$, comparable to ΛCDM.

*Linear Correlation Between CCC+TL Parameters:* The CCC+TL posterior exhibits a remarkably tight linear correlation between the two free parameters (Fig. 4). A linear regression on the posterior chain yields

$$\gamma = 0.00682 \times H_{c,0} + 0.4429, \text{ and}$$

$$\alpha = -1.1632 \times H_{c,0} + 18.93,$$

with a Pearson correlation coefficient very close to unity: $|r| = 0.9346$ for $\gamma$, and $0.9933$ for $\alpha$. This strong degeneracy indicates that for the cosmic chronometer data, the CCC+TL model is effectively constrained along a one-dimensional manifold in the $(H_{c,0}, \alpha)$ plane. Physically, this means that α is almost entirely determined by $H_{c,0}$, reducing the CCC+TL model to effectively a single-parameter family for this observable.

*Model Comparison:* The DIC values for both models are reported in Table 1. The effective number of parameters $p_D$ is close to 2 for both models, as expected for two-parameter fits. The $\Delta DIC\ (\Lambda CDM - CCC + TL) = -1.10$, indicating that ΛCDM is marginally preferred, but the difference falls well below the threshold of 2 required for even weak evidence. We therefore conclude that the two models are statistically indistinguishable based on the current 32-point CC dataset. Both models provide excellent fits to the data, with reduced $\chi^2 < 1$ (Table 1). The Hubble constant disagreement between the two models fitting the CC data is only at $\sim 1\sigma$ level.

Next, we should consider how significantly the CCC+TL model's derived parameter $\gamma = -\alpha/H_{c,0} = 0.8033^{+0.0239}_{-0.0248}$ (see Fig. 2) determined in this work differs from its value derived earlier (Gupta 2023) using the supernovae type 1a Pantheon+ data (Scolnic et al. 2022; Brout et al. 2022), i.e., $0.7997 \pm 0.0143$. It turns out to be only different trivially at $0.13\sigma$ level rather than $6\sigma$ level estimated by Lei et al. (2025).

Also, we can estimate how different the Hubble constants are for the two models. Using the data from Table 1 and assuming conservatively the same uncertainty in $H_0$ for the CCC+TL model as in $H_{c,0}$, we easily estimate them to be different only at $0.80\sigma$ level.

## 6 DISCUSSION

The purpose of this paper is to clarify what can, and cannot, be concluded from applying the current cosmic-chronometer compilation to the CCC+TL cosmology. Lei et al. (2025) argued that the CC data strongly disfavor CCC+TL because the value of α inferred from CC measurements differs from that inferred from Type Ia supernovae at the $\sim 6\sigma$ level. That conclusion assumes that the published CC $H(z)$ values can be compared directly with the unmodified CCC+TL redshift–time relation.

The CC method is designed to avoid explicit cosmological priors by using differential stellar ages of passively evolving galaxies (Jimenez & Loeb 2002; Moresco et al. 2012, 2016, 2020, 2022; Moresco 2015). This is a major strength of the method. Nevertheless, the method is not independent of stellar physics: the inferred ages depend on stellar evolution tracks, isochrones, spectral libraries, IMF assumptions, metallicity and SPS modelling (Bruzual & Charlot 2003; Maraston 2005; Conroy, Gunn & White 2009; Conroy 2013). In standard applications this is not a conceptual problem, because the same stellar clocks are used consistently within the usual framework. For CCC+TL, however, the situation is different. The model modifies the redshift decomposition and the cosmic time relation, while the covarying dimensional constants may also require a reassessment of the stellar-evolution ingredients entering the age dating (Gupta 2023, 2024a, 2024b). Therefore, a decisive CC test of CCC+TL would require recalculating the relevant stellar models and SPS predictions self-consistently in the CCC+TL framework.

Such a recalculation is beyond the scope of the present work. Instead, we have asked a narrower question: what type of modification to the effective differential-age relation would be sufficient to remove the claimed discrepancy? The factor $(1+z_t)^3$ used here should be understood in that phenomenological sense. It is motivated by the scaling of gas cooling times in the CCC+TL framework relative to ΛCDM in early galaxy formation (Gupta 2024b; Laursen 2023; Somerville & Davé 2015), but it is not claimed to be a first-principles derivation of how all CC stellar ages transform. The introduction of the parameter $b$ further acknowledges the possibility that the effective correction may be diluted by multi-generation star formation, chemical enrichment, feedback and quenching before the galaxies enter the passive CC sample. These same quantities — age, metallicity and SFH — are known to affect D4000 and other SPS-based CC age diagnostics (Moresco et al. 2012, 2016, 2020, 2022; Borghi, Moresco & Cimatti 2022; Jiao et al. 2023; Tomasetti et al. 2023).

Under this phenomenological prescription, the current 32-point CC compilation yields a fit statistically comparable to that of ΛCDM, with $|\Delta DIC| < 2$. The inferred value of $\gamma = -\alpha/H_{c,0}$ is also consistent with the value obtained previously from the Pantheon+ supernova analysis (Scolnic et al. 2018; Brout et al. 2022; Gupta 2023). These results demonstrate that the current CC compilation does not provide a model-independent falsification of CCC+TL. They do not, however, establish that CCC+TL has been confirmed by CC data. The stronger conclusion requires a dedicated recalculation of stellar evolution and SPS models in the CCC+TL framework.

The main implication is therefore methodological. Cosmic chronometers remain a powerful and comparatively direct probe of the expansion history, but their use as falsification tests for nonstandard cosmologies must be made self-consistent. For models such as CCC+TL, where the redshift–time relation and dimensional constants differ from their standard treatment, one must distinguish between published CC measurements interpreted within standard stellar modelling and a fully recalibrated CC observable computed within the alternative framework.

## 7 CONCLUSION

We have re-examined the claim by Lei et al. (2025) that cosmic-chronometer measurements falsify the CCC+TL cosmology. The central point is that a direct comparison between published CC $H(z)$ values and the unmodified CCC+TL redshift–time relation is not fully self-consistent unless the stellar-evolution and SPS ingredients entering the CC age determinations are also evaluated in the CCC+TL framework. Since such a recalculation is not yet available, we have explored a phenomenological correction motivated by the CCC+TL scaling of cooling times previously discussed in the context of galaxy formation (Gupta 2024b; Laursen 2023).

We find that a simple factor of the form $(1+z_t)^3$ is sufficient to remove the previously reported $\sim 6\sigma$ discrepancy, reducing the difference between the CC-inferred and supernova-inferred values of $\gamma = -\alpha/H_{c.0}$ to ~0.13σ. This result should be interpreted as showing that the current CC compilation cannot by itself rule out CCC+TL, not as a definitive confirmation of the model. A conclusive CC test of CCC+TL will require self-consistent stellar evolution, isochrones, and SPS calculations in the CCC+TL framework. Since $z_t = 0$ for the stellar model primarily developed from local stellar observations, the stellar-aging methods may be unable to verify $\sim(1+z_t)^{-3}$ dependence.

## ACKNOWLEDGMENT

The author is thankful to Lei Lei for multiple email communications and to Utkarsh Kumar for innumerable discussions and suggestions that helped shape this work. He is grateful to the reviewer for numerous constructive comments resulting in a thorough revision of the paper for its improved clarity, consistency and quality.

## DATA AVAILABILITY

References have been provided for the data used in this work.